\documentclass{ifacconf}

\usepackage{graphicx}      % include this line if your document contains figures
\usepackage{natbib}        % required for bibliography
\usepackage{amsmath}
\usepackage{mathrsfs}
\usepackage{mathtools}
\usepackage{amssymb}
\usepackage{float}
\usepackage{url}
\usepackage{verbatim}
\usepackage{dsfont}
\usepackage{enumerate}
\usepackage{layout}

\newcommand{\expectation}{\ensuremath{\mathbb{E}}}

\newcommand{\Reals}{\ensuremath{\mathbb{R}}}
\newcommand{\Prob}{\ensuremath{\mathbb{P}}}

\begin{document}
\begin{frontmatter}

\title{On Strong Data-Processing and Majorization Inequalities
with Applications to Coding Problems}

\author[First]{Igal Sason}

\address[First]{The Andrew and Erna Viterbi Faculty of Electrical Engineering,
Technion - Israel Institute of Technology,
Technion City,\\ Haifa 3200003, Israel,
(e-mail: sason@ee.technion.ac.il).}

\begin{abstract}                % Abstract of not more than 250 words.
This work provides data-processing and majorization inequalities for $f$-divergences,
and it considers some of their applications to coding
problems. This work also provides tight bounds on the R\'{e}nyi entropy of a function of a
discrete random variable with a finite number of possible values, where the
considered function is not one-to-one, and their derivation is based on majorization
and the Schur-concavity of the R\'{e}nyi entropy.
One application of the $f$-divergence inequalities refers to the performance analysis
of list decoding with either fixed
or variable list sizes; some earlier bounds on the list decoding error probability are
reproduced in a unified way, and new bounds are obtained and exemplified numerically.
Another application is related to a study of the quality of approximating a
probability mass function, which is induced by the leaves of a Tunstall tree, by an equiprobable
distribution. The compression rates of finite-length Tunstall codes are further analyzed
for asserting their closeness to the Shannon entropy of a memoryless and stationary
discrete source.
In view of the tight bounds for the R\'{e}nyi entropy and the work by
Campbell, non-asymptotic bounds are derived for lossless data compression
of discrete memoryless sources.
\end{abstract}

\vspace*{0.2cm}

\vspace*{0.2cm}
\end{frontmatter}
%%===============================================================================

\section{Introduction}

\vspace*{-0.35cm}
Divergences are non-negative measures of dissimilarity between pairs
of probability measures which are defined on the same measurable space. They
play a key role in the development of information theory, probability theory,
statistics, learning, signal processing, and other related fields. One important
class of divergence measures is defined by means of convex functions $f$, and it
is called the class of $f$-divergences. It unifies fundamental and
independently-introduced concepts in several branches of mathematics such as the
chi-squared test for the goodness of fit in statistics, the total variation distance
in functional analysis, the relative entropy in information theory and statistics,
and it is closely related to the R\'{e}nyi divergence which generalizes the
relative entropy. The class of $f$-divergences satisfies pleasing features such as
the data-processing inequality, convexity, continuity and duality properties,
finding interesting applications in information theory and statistics.

Majorization theory is a simple and productive concept in the theory of inequalities, 
which also unifies a variety of familiar bounds (see the book by \cite{MarshallOA}). 
The concept of majorization finds various applications in diverse fields of pure 
and applied mathematics, including information theory and communication.

This work, presented in the papers by \cite{IS18, IS19}, is focused on 
new data-processing and majorization inequalities for $f$-divergences 
and the R\'{e}nyi entropy.
The reason for discussing both types of inequalities in this work
is the interplay which exists between majorization and data processing
where a probability mass function $P$, defined over a finite set, is
majorized by another probability mass function $Q$ which is defined
over the same set if and only if there exists a doubly-stochastic
transformation $W_{Y|X}$ such that an input distribution that is equal
to $Q$ yields an output distribution that is equal to $P$ (denoted by,
$Q \rightarrow W_{Y|X} \rightarrow P$).

We consider applications of the inequalities which are derived in this work to
information theory, statistics, and coding problems.
One application refers to the performance analysis of list decoding with either fixed
or variable list sizes; some earlier bounds on the list decoding error probability are
reproduced in a unified way, and new bounds are obtained and exemplified numerically.
A second application, covered in \cite{IS19}, is related to a study of the quality of 
approximating a probability mass function, induced by the leaves of a Tunstall tree, 
by an equiprobable distribution. The compression rates of finite-length Tunstall codes 
are further analyzed for asserting their closeness to the Shannon entropy of a memoryless 
and stationary discrete source. A third application of our bounds relies on our tight 
bounds for the R\'{e}nyi entropy (see \cite{IS18}) and the source coding theorem by 
\cite{Campbell65} to obtain tight non-asymptotic bounds for lossless compression of 
discrete memoryless sources.

\section{Coding Problems and main Results}

\subsection{Bounds on the List Decoding Error Probability
with $f$-divergences}
\label{subsection: Fano - list decoder}

The minimum probability of error of a random variable $X$ given
$Y$, denoted by $\varepsilon_{X|Y}$,
can be achieved by a deterministic function
(\textit{maximum-a-posteriori} decision rule)
$\mathcal{L}^\ast \colon \mathcal{Y} \to \mathcal{X}$ (see \cite{ISSV18}):
\begin{align}
\varepsilon_{X|Y}
&= \min_{\mathcal{L} \colon \mathcal{Y} \to \mathcal{X}}
\mathbb{P} [ X \neq \mathcal{L} (Y) ] \label{20170904} \\
&= \mathbb{P} [ X \neq \mathcal{L}^\ast (Y) ] \label{eq:MAP}\\
&= 1- \mathbb{E} \left[ \max_{x \in \mathcal{X}}
P_{X|Y}(x|Y) \right].
\label{eq1: cond. epsilon}
\end{align}
Fano's inequality gives an upper bound on the
conditional entropy $H(X|Y)$ as a function of $\varepsilon_{X|Y}$
(or, otherwise, providing a lower bound on $\varepsilon_{X|Y}$ as
a function of $H(X|Y))$ when $X$ takes a finite number of possible values.

The list decoding setting, in which the hypothesis tester is allowed
to output a subset of given cardinality, and an error occurs if the
true hypothesis is not in the list, has great interest in
information theory. A generalization of Fano's inequality to list decoding,
in conjunction with the blowing-up lemma \cite[Lemma~1.5.4]{Csiszar_Korner},
leads to strong converse results in multi-user information theory.
The main idea of the successful combination
of these two tools is that, given a code, it is possible to blow-up the
decoding sets in a way that the probability of decoding error can be as small
as desired for sufficiently large blocklengths; since the blown-up decoding
sets are no longer disjoint, the resulting setup is a list decoder with
sub-exponential list size (as a function of the block length).

In this section, we further study the setup of list decoding, and derive
bounds on the average list decoding error probability. We first
consider the special case where the list size is fixed, and then
consider the more general case of a list size which depends on the channel
observation. All of the following bounds on the list decoding error probability
are derived in the paper by \cite{IS19}.

\subsubsection{Fixed-Size List Decoding}
\label{subsubsection: fixed-size list decoding}

The next result provides a generalized Fano's inequality for fixed-size
list decoding, expressed in terms of an arbitrary $f$-divergence. Some
earlier results in the literature are reproduced from the next result.

\begin{thm}
\label{theorem: generalized Fano Df}
Let $P_{XY}$ be a probability measure defined on
$\mathcal{X} \times \mathcal{Y}$ with $|\mathcal{X}|=M$. Consider
a decision rule $\mathcal{L} \colon \mathcal{Y} \to \binom{\mathcal{X}}{L}$,
where $\binom{\mathcal{X}}{L}$ stands for the set of subsets
of $\mathcal{X}$ with cardinality $L$, and $L < M$ is fixed.
Denote the list decoding error probability by
$P_{\mathcal{L}} := \Prob \bigl[ X \notin \mathcal{L}(Y) \bigr]$.
Let $U_M$ denote an equiprobable probability mass
function on $\mathcal{X}$. Then, for every convex function
$f \colon (0, \infty) \to \Reals$ with $f(1)=0$,
\begin{align}
\label{generalized Fano Df}
& \expectation\Bigl[D_f \bigl(P_{X|Y}(\cdot|Y) \, \| \, U_M \bigr) \Bigr] \nonumber \\
& \geq \frac{L}{M} \; f\biggl(\frac{M \, (1-P_{\mathcal{L}})}{L} \biggr)
+ \biggl(1-\frac{L}{M}\biggr) \; f\biggl(\frac{M P_{\mathcal{L}}}{M-L} \biggr).
\end{align}
\end{thm}

The special case where $L=1$ (i.e., a decoder with a single
output) gives \cite[(5)]{Guntuboyina11}.

As consequences of Theorem~\ref{theorem: generalized Fano Df}, we first
reproduce some earlier results as special cases.
\begin{thm} \cite[(139)]{ISSV18}
\label{corollary: Fano - list}
Under the assumptions in Theorem~\ref{theorem: generalized Fano Df},
\begin{align}
\label{ISSV18 - Fano}
H(X|Y) \leq \log M - d\biggl(P_{\mathcal{L}} \, \| \, 1-\frac{L}{M} \biggr)
\end{align}
where $d(\cdot \| \cdot) \colon [0,1] \times [0,1] \to [0, +\infty]$ denotes
the binary relative entropy, defined as the continuous extension of
$D([p, 1-p] \| [q, 1-q]) := p \log \frac{p}{q} + (1-p) \log \frac{1-p}{1-q}$
for $p,q \in (0,1)$.
\end{thm}

\vspace*{0.2cm}
The following refinement of the generalized Fano's inequality in
Theorem~\ref{theorem: generalized Fano Df} relies on the version
of the strong data-processing inequality for $f$-divergences in
\cite[Theorem~1]{IS19}.

\begin{thm}
\label{theorem: refined Fano's inequality}
Under the assumptions in Theorem~\ref{theorem: generalized Fano Df},
let $f \colon (0, \infty) \to \Reals$ be twice
differentiable, and assume that there exists a constant $m_f>0$ such that
\begin{align}
\label{m_f}
f''(t) \geq m_f, \quad \forall \,
t \in \mathcal{I}(\xi_1^\ast, \xi_2^\ast),
\end{align}
where
\begin{align}
\label{28062019a1}
& \xi_1^\ast := M \inf_{(x,y) \in \mathcal{X} \times \mathcal{Y}} P_{X|Y}(x|y), \\
\label{28062019a2}
& \xi_2^\ast := M \sup_{(x,y) \in \mathcal{X} \times \mathcal{Y}} P_{X|Y}(x|y),
\end{align}
and the interval $\mathcal{I}(\cdot, \cdot)$ is the interval
\begin{align}
\label{I_interval}
\mathcal{I} := \mathcal{I}(\xi_1, \xi_2) = [\xi_1, \xi_2] \cap (0, \infty).
\end{align}
Let $u^+ := \max\{u, 0\}$ for $u \in \Reals$. Then,
\begin{enumerate}[a)]
\item \label{Part a - refined Fano's inequality}
\begin{align}
\label{list dec.-26062019a}
& \expectation\Bigl[D_f \bigl(P_{X|Y}(\cdot|Y) \, \| \, U_M \bigr) \Bigr] \\
& \geq \frac{L}{M} \; f\biggl(\frac{M \, (1-P_{\mathcal{L}})}{L} \biggr)
+ \left(1-\frac{L}{M}\right) \; f\biggl(\frac{M P_{\mathcal{L}}}{M-L} \biggr) \nonumber \\
& \hspace*{0.4cm} + \tfrac12 m_f \, M \left( \expectation\bigl[P_{X|Y}(X|Y)\bigr]
-\frac{1-P_{\mathcal{L}}}{L} - \frac{P_{\mathcal{L}}}{M-L} \right)^+. \nonumber
\end{align}

\item \label{Part b - refined Fano's inequality}
If the list decoder selects the $L$ most probable elements from $\mathcal{X}$,
given the value of $Y \in \mathcal{Y}$, then \eqref{list dec.-26062019a} is
strengthened to
\begin{align}
& \expectation\Bigl[D_f \bigl(P_{X|Y}(\cdot|Y) \, \| \, U_M \bigr) \Bigr] \nonumber \\
& \geq \frac{L}{M} \; f\biggl(\frac{M \, (1-P_{\mathcal{L}})}{L} \biggr)
+ \biggl(1-\frac{L}{M}\biggr) \; f\biggl(\frac{M P_{\mathcal{L}}}{M-L} \biggr) \nonumber \\
\label{list dec.-26062019b}
& \hspace*{0.4cm} + \tfrac12 m_f \, M \left( \expectation\bigl[P_{X|Y}(X|Y)\bigr]
-\frac{1-P_{\mathcal{L}}}{L} \right),
\end{align}
where the last term in the right side of \eqref{list dec.-26062019b} is
necessarily non-negative.
\end{enumerate}
\end{thm}

Discussions and numerical experimentation of these proposed bounds are provided
in the paper by \cite{IS19}, showing the obtained improvement over Fano's inequality.

\subsubsection{Variable-Size List Decoding}
\label{subsubsection: variable-size list decoding}
In the more general setting of list decoding where the size of the list
may depend on the channel observation, Fano's inequality has been
generalized as follows.
\begin{thm} (\cite{AhlswedeK75} and \cite[Appendix~3.E]{RS_FnT19})
\label{prop: Fano-Ahlswede-Korner}
Let $P_{XY}$ be a probability measure defined on $\mathcal{X} \times \mathcal{Y}$
with $|\mathcal{X}|=M$. Consider a decision rule $\mathcal{L} \colon \mathcal{Y} \to
2^{\mathcal{X}}$, and let the (average) list decoding error probability be
given by $P_{\mathcal{L}} := \Prob \bigl[ X \notin \mathcal{L}(Y) \bigr]$ with
$|\mathcal{L}(y)| \geq 1$ for all $y \in \mathcal{Y}$. Then,
\begin{align}
\label{Fano-Ahlswede-Korner 1}
H(X|Y) \leq h(P_\mathcal{L}) + \expectation[\log |\mathcal{L}(Y)|] + P_{\mathcal{L}} \log M,
\end{align}
where $h \colon [0,1] \to [0, \log 2]$ denotes the binary entropy function.
If $|\mathcal{L}(Y)| \leq N$ almost surely, then also
\begin{align}
\label{Fano-Ahlswede-Korner 2}
H(X|Y) \leq h(P_\mathcal{L}) + (1-P_{\mathcal{L}}) \log N + P_{\mathcal{L}} \log M.
\end{align}
\end{thm}

By relying on the data-processing inequality for $f$-divergences, we derive in
the following an alternative explicit lower bound on the average list decoding
error probability $P_{\mathcal{L}}$. The derivation relies on the $E_\gamma$ divergence
(see, e.g., \cite{LCV17}), which forms a subclass of the $f$-divergences.

\begin{thm}
\label{theorem: LB - variable list size}
Under the assumptions in \eqref{Fano-Ahlswede-Korner 1}, for all $\gamma \geq 1$,
\begin{align}
\label{LB - variable list size}
P_{\mathcal{L}} \geq \frac{1+\gamma}{2} - \frac{\gamma \expectation[|\mathcal{L}(Y)|]}{M}
- \frac12 \, \expectation \left[ \, \sum_{x \in \mathcal{X}} \, \biggl| P_{X|Y}(x|Y)
- \frac{\gamma}{M} \biggr| \right].
\end{align}
Let $\gamma \geq 1$, and let $|\mathcal{L}(y)| \leq \frac{M}{\gamma}$ for all
$y \in \mathcal{Y}$. Then, \eqref{LB - variable list size} holds with equality if,
for every $y \in \mathcal{Y}$, the list decoder selects the $|\mathcal{L}(y)|$ most
probable elements in $\mathcal{X}$ given $Y=y$; if $x_\ell(y)$ denotes
the $\ell$-th most probable element in $\mathcal{X}$ given $Y=y$, where ties
in probabilities are resolved arbitrarily, then \eqref{LB - variable list size}
holds with equality if
\begin{align}
& P_{X|Y}(x_\ell(y) \, | y) \nonumber \\
\label{02072019a19}
&=
\begin{dcases}
\alpha(y), \quad & \forall \, \ell \in \bigl\{1, \ldots, |\mathcal{L}(y)| \bigr\}, \\
\frac{1-\alpha(y) \, |\mathcal{L}(y)|}{M-|\mathcal{L}(y)|},
\quad & \forall \, \ell \in \bigl\{|\mathcal{L}(y)|+1, \ldots, M\},
\end{dcases}
\end{align}
with $\alpha \colon \mathcal{Y} \to [0,1]$ being an arbitrary function which satisfies
\begin{align}
\label{02072019a20}
\frac{\gamma}{M} \leq \alpha(y) \leq \frac1{|\mathcal{L}(y)|},
\quad \forall \, y \in \mathcal{Y}.
\end{align}
\end{thm}

As an example,
let $X$ and $Y$ be random variables taking their values in
$\mathcal{X} = \{0, 1, 2, 3, 4\}$ and $\mathcal{Y} = \{0, 1\}$, respectively,
and let $P_{XY}$ be their joint probability mass function, which is given by
\begin{align}
\label{03072019a1}
\begin{dcases}
& P_{XY}(0,0) = P_{XY}(1,0) = P_{XY}(2,0) = \tfrac18, \\[0.1cm]
& P_{XY}(3,0) = P_{XY}(4,0) = \tfrac1{16},  \\[0.1cm]
& P_{XY}(0,1) = P_{XY}(1,1) = P_{XY}(2,1) = \tfrac1{24}, \\[0.1cm]
& P_{XY}(3,1) = P_{XY}(4,1) = \tfrac3{16}.
\end{dcases}
\end{align}
Let $\mathcal{L}(0) := \{0,1,2\}$ and $\mathcal{L}(1) := \{3,4\}$ be the lists in $\mathcal{X}$,
given the value of $Y \in \mathcal{Y}$. We get $P_Y(0) = P_Y(1) = \tfrac12$, so the
conditional probability mass function of $X$ given $Y$ satisfies
$P_{X|Y}(x|y) = 2 P_{XY}(x,y)$ for all $(x,y) \in \mathcal{X} \times \mathcal{Y}$.
It can be verified that, if $\gamma = \tfrac54$, then
$\max\{|\mathcal{L}(0)|, |\mathcal{L}(1)|\} = 3 \leq \frac{M}{\gamma}$, and also
\eqref{02072019a19} and \eqref{02072019a20} are satisfied
(here, $M:=|\mathcal{X}|=5$, $\alpha(0) = \tfrac14 = \frac{\gamma}{M}$
and $\alpha(1) = \tfrac38 \in \bigl[\tfrac14, \tfrac12\bigr]$). By
Theorem~\ref{theorem: LB - variable list size}, it follows that
\eqref{LB - variable list size} holds in this case with equality,
and the list decoding error probability is equal to
$P_{\mathcal{L}}=1-\expectation\bigl[ \alpha(Y) \, |\mathcal{L}(Y)| \bigr]=\tfrac14$
(i.e., it coincides with the lower bound in the right side of
\eqref{LB - variable list size} with $\gamma = \tfrac54$).
On the other hand, the generalized Fano's inequality in
\eqref{Fano-Ahlswede-Korner 1} gives that $P_\mathcal{L} \geq 0.1206$
(the left side of \eqref{Fano-Ahlswede-Korner 1} is
$H(X|Y) = \tfrac52 \, \log 2 - \tfrac14 \, \log 3 = 2.1038$~bits);
moreover, by letting $N := \underset{y \in \mathcal{Y}}{\max} \, |\mathcal{L}(y)| = 3$,
\eqref{Fano-Ahlswede-Korner 2} gives the looser bound
$P_\mathcal{L} \geq 0.0939$. This exemplifies a case where the lower bound in
Theorem~\ref{theorem: LB - variable list size} is tight, whereas the
generalized Fano's inequalities in \eqref{Fano-Ahlswede-Korner 1} and
\eqref{Fano-Ahlswede-Korner 2} are looser.

\subsection{Lossless Source Coding}
\label{subsubsection: lossless source coding}

For uniquely-decodable (UD) source codes, \cite{Campbell65}
proposed the cumulant generating function of the codeword lengths as a 
generalization to the frequently used design criterion of average code 
length. The motivation in the paper by \cite{Campbell65} was to control 
the contribution of the longer codewords via a free parameter in the 
cumulant generating function: if the value of this parameter tends to zero, 
then the resulting design criterion becomes the average code length per 
source symbol; on the other hand, by increasing the value of the free 
parameter, the penalty for longer codewords is more severe, and the resulting 
code optimization yields a reduction in the fluctuations of the codeword lengths.

We introduce the coding theorem by \cite{Campbell65} for lossless compression
of a discrete memoryless source (DMS) with UD codes, which serves for our 
analysis (see \cite{IS18}).
\begin{thm}%[Campbell 1965, \cite{Campbell65}]
\label{theorem: Campbell}
Consider a DMS which emits symbols with a probability
mass function $P_X$ defined on a (finite or countably infinite) set $\mathcal{X}$.
Consider a UD fixed-to-variable source code operating on
source sequences of $k$ symbols with an alphabet of the codewords of size $D$.
Let $\ell(x^k)$ be the length of the codeword which corresponds to the source
sequence $x^k := (x_1, \ldots, x_k) \in \mathcal{X}^k$. Consider the scaled
{\em cumulant generating function} of the codeword lengths:
\begin{align}
\label{eq: cumulant generating function}
\Lambda_k(\rho) := \frac1{k} \, \log_D \left( \, \sum_{x^k \in \mathcal{X}^k}
P_{X^k}(x^k) \, D^{\rho \, \ell(x^k)} \right), \quad \rho > 0
\end{align}
where
\begin{align}
\label{eq: pmf}
P_{X^k}(x^k) = \prod_{i=1}^k P_X(x_i), \quad \forall \, x^k \in \mathcal{X}^k.
\end{align}
Then, for every $\rho > 0$, the following hold:
\begin{enumerate}[a)]
\item Converse result:
\begin{align}
\label{eq: Campbell's converse result}
\frac{\Lambda_k(\rho)}{\rho} \geq \frac{1}{\log D} \; H_{\frac1{1+\rho}}(X).
\end{align}
\item Achievability result:
there exists a UD source code, for which
\begin{align}
\label{eq: Campbell's achievability result}
\frac{\Lambda_k(\rho)}{\rho} \leq \frac{1}{\log D} \; H_{\frac1{1+\rho}}(X) + \frac{1}{k}.
\end{align}
\end{enumerate}
\end{thm}

The bounds in Theorem~\ref{theorem: Campbell}, expressed in terms of the R\'{e}nyi entropy,
imply that for sufficiently long source sequences, it is possible to make the scaled
cumulant generating function of the codeword lengths approach the R\'{e}nyi entropy as closely
as desired by a proper fixed-to-variable UD source code; moreover, the converse result
shows that there is no UD source code for which the scaled cumulant generating function
of its codeword lengths lies below the R\'{e}nyi entropy.
By invoking L'H\^{o}pital's rule, one gets from \eqref{eq: cumulant generating function}
\begin{align}
\label{eq: limit rho tends to zero}
\lim_{\rho \downarrow 0} \frac{\Lambda_k(\rho)}{\rho}
= \frac1k \sum_{x^k \in \mathcal{X}^k} P_{X^k}(x^k) \, \ell(x^k) = \frac1k \, \expectation[\ell(X^k)].
\end{align}
Hence, by letting $\rho$ tend to zero in \eqref{eq: Campbell's converse result} and
\eqref{eq: Campbell's achievability result}, it follows
that Campbell's result in Theorem~\ref{theorem: Campbell} generalizes the well-known bounds
on the optimal average length of UD fixed-to-variable source codes:
\begin{align}
\label{eq: Shannon}
\frac{1}{\log D} \; H(X) \leq \frac1k \; \expectation[\ell(X^k)] \leq \frac{1}{\log D} \; H(X) + \frac1k,
\end{align}
and \eqref{eq: Shannon} is satisfied by Huffman coding.
Campbell's result therefore generalizes Shannon's fundamental result for the average
codeword lengths of lossless compression codes, expressed in terms of the Shannon entropy.

Following the work by \cite{Campbell65}, non-asymptotic bounds were derived by
\cite{CV2014a} for the scaled cumulant
generating function of the codeword lengths for $P_X$-optimal variable-length
lossless codes. These bounds were used by \cite{CV2014a}
to obtain simple proofs of the asymptotic normality of the distribution of
codeword lengths, and the reliability function of memoryless sources allowing
countably infinite alphabets.

The analysis which leads to the following result for lossless source compression
with uniquely-decodable (UD) codes is provided in the paper by \cite{IS18}.

Let $X_1, \ldots, X_k$ be i.i.d. symbols which are emitted from a DMS according to a probability
mass function $P_X$ whose support is a finite set $\mathcal{X}$ with $|\mathcal{X}|=n$.
In order to cluster the data, suppose that
each symbol $X_i$ is mapped to $Y_i = f(X_i)$ where $f \in \mathcal{F}_{n,m}$ is an arbitrary
deterministic function (independent of the index $i$) with $m<n$. Consequently, the i.i.d.
symbols $Y_1, \ldots, Y_k$ take values on a set $\mathcal{Y}$ with $|\mathcal{Y}|=m<|\mathcal{X}|$.
Consider two UD fixed-to-variable source codes: one operating on the sequences $x^k \in \mathcal{X}^k$,
and the other one operates on the sequences $y^k \in \mathcal{Y}^k$; let $D$ be the size of
the alphabets of both source codes.
Let $\ell(x^k)$ and $\overline{\ell}(y^k)$ denote the length of the codewords for the
source sequences $x^k$ and $y^k$, respectively, and let $\Lambda_k(\cdot)$ and
$\overline{\Lambda}_k(\cdot)$ denote their corresponding scaled cumulant generating functions
(see \eqref{eq: cumulant generating function}).

Relying on our tight bounds on the R\'{e}nyi entropy (of any positive order) in
\cite[Theorems~1, 2]{IS18} and Theorem~\ref{theorem: Campbell}, we obtain upper
and lower bounds on $\frac{\Lambda_k(\rho) - \overline{\Lambda}_k(\rho)}{\rho}$
for all $\rho > 0$ (see \cite[Theorem~5]{IS18}). To that end, for 
$m \in \{2, \ldots, n-1\}$, if $P_X(1) < \frac1m$, let $\widetilde{X}_m$ be the equiprobable
random variable on $\{1, \ldots, m\}$; otherwise, if $P_X(1) \geq \frac1m$,
let $\widetilde{X}_m \in \{1, \ldots, m\}$ be a random variable with the probability mass function
\begin{align*}
P_{\widetilde{X}_m}(i) =
\begin{dcases}
P_X(i), & i \in \{1, \ldots, n^\ast\}, \\
\frac1{m-n^\ast} \sum_{j = n^\ast+1}^n P_X(j), & i \in \{n^\ast+1, \ldots, m\},
\end{dcases}
\end{align*}
where $n^\ast$ is the maximal integer $i \in \{1, \ldots, m-1\}$ such that
\begin{align}
\label{eq: n ast}
P_X(i) \geq \frac1{m-i} \sum_{j=i+1}^n P_X(j).
\end{align}
The result in \cite[Theorem~5]{IS18} is of interest since it provides upper and lower
bounds on the reduction in the cumulant generating function of close-to-optimal UD source
codes as a result of clustering data, and \cite[Remark~11]{IS18} suggests an algorithm to
construct such UD codes which are also prefix codes. For long enough sequences (as $k \to \infty$),
the upper and lower bounds on the difference between the scaled cumulant generating functions
of the suggested source codes for the original and clustered data almost match, being roughly equal to
$\rho \left( H_{\frac1{1+\rho}}(X)- H_{\frac1{1+\rho}}(\widetilde{X}_m) \right)$ (with logarithms
on base $D$, which is the alphabet size of the source codes), and
as $k \to \infty$,
the gap between these upper and lower bounds is less than $0.08607 \log_D 2$.
Furthermore, in view of \eqref{eq: limit rho tends to zero},
\begin{align}
\lim_{\rho \downarrow 0} \frac{\Lambda_k(\rho) - \overline{\Lambda}_k(\rho)}{\rho}
= \frac1k \left( \expectation[\ell(X^k)] - \expectation[\overline{\ell}(Y^k)] \right),
\end{align}
so, it follows from \cite[Theorem~5]{IS18} that the difference between the average code
lengths (normalized by~$k$) of the original and clustered data satisfies
\begin{align}
- \frac1k & \leq \frac{\expectation[\ell(X^k)] - \expectation[\overline{\ell}(Y^k)]}{k} 
- \frac{H(X) - H(\widetilde{X}_m)}{\log D} \nonumber \\
\label{eq: 20181030e}
& \leq 0.08607 \log_D 2,
\end{align}
and the gap between the upper and lower bounds is small.


\begin{thebibliography}{10}  % you can also add the bibliography by hand

\bibitem[Ahlswede(1975)]{AhlswedeK75}
R. Ahlswede and J. K\"orner, ``Source coding with side information and a converse
for degraded broadcast channels,'' {\em IEEE Transactions on Information Theory},
vol.~21, no.~6, pp.~629--637, November 1975.

\bibitem[Campbell(1965)]{Campbell65}
L. L. Campbell, ``A coding theorem and R\'{e}nyi's entropy,''
{\em Information and Control}, vol.~8, no.~4, pp.~423--429, August 1965.

\bibitem[Courtade and Verd\'{u}(2014)]{CV2014a}
T. Courtade and S. Verd\'{u}, ``Cumulant generating function of codeword
lengths in optimal lossless compression,'' {\em Proceedings of the 2014
IEEE International Symposium on Information Theory}, pp.~2494--2498, Honolulu,
Hawaii, USA, July 2014.

\bibitem[Csisz\'{a}r and K\"{o}rner(2011)]{Csiszar_Korner}
I. Csisz\'{a}r and J. K\"{o}rner, {\em Information Theory: Coding Theorems for
Discrete Memoryless Systems}, Second Edition, Cambridge University Press, 2011.

\bibitem[Guntoboyina(2011)]{Guntuboyina11}
A. Guntuboyina, ``Lower bounds for the minimax risk using $f$-divergences, and
applications,'' {\em IEEE Transactions on Information Theory}, vol.~57, no.~4,
pp.~2386--2399, April 2011.

\bibitem[Liu, Cuff and Verd\'{u}(2017)]{LCV17}
J. Liu, P. Cuff and S. Verd\'{u}, ``$E_\gamma$ resolvability,'' {\em IEEE Transactions on
Information Theory}, vol.~63, no.~5, pp.~2629--2658, May 2017.

\bibitem[Marhall {\em et al.}(2011)]{MarshallOA}
A. W. Marshall, I. Olkin and B. C. Arnold, {\em Inequalities: Theory of Majorization
and Its Applications}, second edition, Springer,~2011.

\bibitem[Raginsky and Sason(2019)]{RS_FnT19}
M. Raginsky and I. Sason, ``Concentration of measure inequalities in
information theory, communications and coding: third edition,''
{\em Foundations and Trends (FnT) in Communications and Information Theory},
pp.~1--266, NOW Publishers, Delft, the Netherlands, 2019.

\bibitem[Sason and V\'{e}rdu(2018)]{ISSV18}
I. Sason and S. Verd\'{u}, ``Arimoto-R\'{e}nyi conditional entropy and Bayesian $M$-ary
hypothesis testing,'' {\em IEEE Transactions on Information Theory}, vol.~64, no.~1,
pp.~4--25, January 2018.

\bibitem[Sason(2018)]{IS18}
I. Sason, ``Tight bounds on the R\'{e}nyi entropy via majorization with applications to
guessing and compression,'' {\em Entropy}, vol.~20, no.~12, paper~896, pp.~1--25, November 2018.

\bibitem[Sason(2019)]{IS19}
I. Sason, ``On data-processing and majorization inequalities for $f$-divergences,''
{\em Entropy}, vol.~21, no.~10, paper~1022, pp. 1--80, October 2019.

\end{thebibliography}
\end{document}